# Condensation of Self-assembled Lyotropic Chromonic Liquid Crystal Sunset Yellow in Aqueous Solutions Crowded with Polyethylene glycol and Doped with Salt

*Heung-Shik Park[a], Shin-Woong Kang[b], Luana Tortora[a],*

*Satyendra Kumar[c], and Oleg D. Lavrentovich*[,a]

[a] Liquid Crystal Institute and Chemical Physics Interdisciplinary Program,
Kent State University, Kent, OH 44242,

[b] Department of BIN Fusion Technology, Chonbuk National University, Jeonju, Korea 561-756,

[c] Department of Physics, Kent State University, Kent, OH 44242

We use optical and fluorescence microscopy, densitometry, cryo-transmission electron microscopy (cryo-TEM), spectroscopy, and synchrotron X-ray scattering, to study the phase behavior of the reversible self-assembled chromonic aggregates of an anionic dye Sunset Yellow (SSY) in aqueous solutions crowded with an electrically neutral polymer polyethylene glycol (PEG) and doped with the salt NaCl. PEG causes the isotropic SSY solutions to condense into a liquid-crystalline region with a high concentration of SSY aggregates, coexisting with a PEG-rich isotropic (I) region. PEG added to the homogeneous nematic (N) phase causes separation into the coexisting N and I domains; the SSY concentration in the N domains is higher than the original concentration of PEG-free N phase. Finally, addition of PEG to the highly concentrated homogeneous N phase causes separation into the coexisting columnar hexagonal (C) phase and I phase. This behavior can be qualitatively explained by the depletion (excluded volume) effects that act at two different levels: at the level of aggregate assembly from monomers and short aggregates and at the level of inter-aggregate packing. We also show a strong effect of a monovalent salt NaCl on phase diagrams that is different for high and low concentrations of SSY. Upon the addition of salt, dilute I solutions of SSY show appearance of the condensed N domains, but the highly concentrated C phase transforms into a coexisting I and N domains. We suggest that the salt-induced screening of electric charges at the surface of chromonic aggregates leads to two different effects: (a) increase of the scission energy and the contour length of aggregates, and (b) decrease of the persistence length of SSY aggregates.

## 1. Introduction

Lyotropic chromonic liquid crystals (LCLCs) form through the process of reversible self-assembly and represent a broad but not well understood class of soft matter [1-3]. The LCLC molecules have a plank-like or disk-like polyaromatic central core with polar groups at the periphery. In water, the molecules stack on top of each other forming the so-called H-aggregates, leaving the polar solubilizing groups at the aggregate-water interface [1-3]. A typical separation between the adjacent molecules along the stacking direction is about (0.33-0.34) nm. When the polar groups are fully ionized, the line density of electric charge along the aggregate can be very high, e.g., $\sim 6e/nm$ ($e$ is the electron's charge) for the LCLC molecules with two ionic groups. The stacking distance and the line charge make LCLC aggregates similar to the double-strand B-DNA molecules. The important difference is that in LCLCs, there are no chemical bonds to fix the length of aggregates. As the concentration of a chromonic material increases, the aggregates multiply, elongate, and align parallel to the common direction, called the director $\hat{\mathbf{n}}$. The two most commonly met phases are the uniaxial nematic (N) phase and the columnar (C) hexagonal phase [1-3]. In earlier publications, one often finds the term "M phase", where "M" stands for "middle." We use the term "C phase" instead of "M phase" whenever the presence of a hexagonal columnar phase is established by a structural study, such as an X-ray diffraction performed in this work. The aggregates' length depends not only on the concentration, but also on the specific details of molecular interactions, temperature [1, 2, 4], ionic content [5-7], pH of the solution [7], type of the side groups [2, 3]. The LCLCs thus represent an interesting self-assembled system with an orientational and positional order that is highly sensitive to a number of factors. The chromonic family should be extended to include the DNA nucleotides, such as guanosine derivatives [8, 9]. LCLCs show a potential for new applications, such as controlled drug delivery [1], biosensing [10], preparation of optically anisotropic films [11], micro-patterning [12], nano-fabrication [13], etc.

One of the most intriguing questions arising in the studies of LCLC is how their self-assembly is affected by additives, either charged, such as salts [5, 7, 14-16], or non-charged, such as neutral polymers [17, 18]. The two main mechanisms associated with the role of additives are (a) electrostatic interactions within and between the aggregates and (b) excluded volume effects. Neither of the two is well understood.

Exploration of the electrostatic effects started with the observation made by Yu and Saupe [5] that the addition of a salt NaCl to one of the most studied LCLC materials, disodium cromoglycate (DSCG), increases the temperature $T_{N \rightarrow N+I}$ at which the homogeneous N phase transforms into the biphasic N-isotropic (I) coexistence region, as well as the temperature $T_{N+I \rightarrow I}$ of the complete melting. Kostko et al. [14] supported this conclusion of the salt-induced increase of $T_{N \rightarrow N+I}$ and $T_{N+I \rightarrow I}$ for the case of salts with small cations, Na$^+$, K$^+$, but found also an opposite effect of destabilization of the N phase by salts with large organic cations, such as tetraethylammonium bromide and tetrabutylammonium bromide. Prasad et al. [15] showed that NaCl increases the viscosity of another LCLC material, the disodium salt of 6-hydroxy-5-[(4-sulfophenyl)azo]-2-naphthalenesulfonic acid, also known as Sunset Yellow FCF (SSY), but did not find any change in the transition temperatures. In the subsequent works [7, 16], the effect of NaCl on the transition temperatures of SSY was observed, but the two reported trends were exactly opposite: our group observed an increase in $T_{N \to N+I}$ and $T_{N+I \to I}$ [7], while Jones et al. [16] showed that these temperatures decrease upon the addition of NaCl; as clarified in what follows, both trends are possible and their prevalence depends on the initial concentration of SSY

The depletion effects in LCLCs are not much clearer. In the classic picture, the excluded volume effects are considered for colloidal dispersions of particles with a fixed shape, say, solid rods of constant length and diameter. In this picture, an added depletion agent, such as neutral spheres, forces the rods to pack more closely, as the spheres cannot penetrate the rods and the volume available for them is maximized when the two species are separated [19, 20]. One of the most efficient depletion agents is the neutral non-adsorbing polymer polyethylene glycol (PEG). In water, a PEG molecule behaves as a random coil that can be approximated by a sphere of a certain radius of gyration $r_g$. PEG-induced "depletion attraction" has been observed for a variety of systems, including the solutions of DNA [21-24]. For LCLCs, the data are scarce. Simon et al. [17] demonstrated that some water-soluble polymers added to isotropic DSCG solutions cause the formation of birefringent droplets, with polymer creating a shell around the LCLC droplet. PEG of a molecular weight 600-1,500 was reported as producing no effect on LCLCs [17]. In contrast, PEG of a molecular weight 3,350 was shown to condense the solutions of DSCG into orientationally and positionally ordered domains of various morphologies, such as tactoids of the N phase, and toroids of the C phase [18]. We expect that the depletion effects in LCLCs are more complex than in dispersions with particles of fixed shape and length. Since the LCLC assembly is non-covalent, the neutral additives can influence the system at two different levels: at the level of aggregate assembly from smaller aggregates and monomers, and at the level of inter-aggregate interaction.

Motivated by these considerations, in the present work, we explore the complete phase diagram of the three-component system formed by SSY, PEG and water, as well as formulations with an added monovalent salt NaCl. The aggregate structure of SSY is somewhat better established as compared to that of DSCG. The stacks of SSY in water contain only one molecule in cross-section, the plane of which is perpendicular to the aggregate axis [25-28]. We find that PEG condenses SSY into ordered LC domains coexisting with a PEG-rich I phase. In the condensed LC region, the distance between the SSY aggregates decreases and the overall concentration of SSY increases (as compared to the parental PEG-free solution) as the concentration of PEG increases. Addition of NaCl can either enhance the condensing effect of PEG, or suppress it, depending on the concentrations. We also find that the orientation of SSY aggregates at the LC-I interface of small (submicron) LC domains can be tangential, normal or tilted. The behavior is different from the behavior of large supramicron LC domains in which the aggregates align tangentially to the LC-I interface [18]. The observation suggests a finite strength of the surface anchoring, similar to the recent findings in Ref. [29].

## 2. Experimental Techniques

**Materials**. Sunset Yellow FCF (SSY) of a purity of 95.7% was purchased from Sigma Aldrich (St. Louis, MO) and purified by the procedure established earlier [7, 25]. The purified SSY was dehydrated by placing it in a vacuum oven for two days before use, because dried SSY is easily hydrated during the storage [7]. PEG (molecular weight 3,350) whose radius of gyration is $r_g \approx 2$ nm [30], was purchased from Sigma Aldrich and used without further purification. Fluorescein isothiocyanate PEG (FITC-PEG) with a molecular weight 3,400 was purchased from Nanocs Inc. (New York, NY). Distilled water further purified with a Millipore water purification system (resistivity $\geq 18.1$ M$\Omega$·cm) was used to prepare all the solutions. In most cases, we use weight % units for the concentration $c_{SSY}$ of SSY, $c_{PEG}$ of PEG and $c_w$ of water, so that in all the mixtures, $c_{SSY} + c_{PEG} + c_w = 100$ wt.%.

**Phase diagram study.** The phase identification was performed using polarizing optical microscopy (POM) and X-ray diffraction measurements. For optical observations, the samples were prepared by placing a drop of the mixture between two glass plates separated by mylar spacers (thickness 12 μm) and sealing the edges with an epoxy glue (Davcon).

**Fluorescence microscopy.** To get an insight into partitioning the components in phase separated mixtures, we added a small amount (1~2wt.%) of FITC-PEG to its non-fluorescent counterpart PEG. We used the Olympus Fluoview confocal microscope BX50 with an Ar laser ($\lambda = 488$ nm) for the excitation of FITC-PEG. The fluorescence was detected in the spectral range 510-550 nm.

**Cryo-TEM.** To prepare the vitrified sample for cryo-TEM, a 5 μl of mixture was dropped on a holey carbon grid (Ted Pella, Redding, CA) in a controlled environment vitrification chamber (CEVS, Vitrobot, FEI) in which the atmosphere surrounding the sample grid was kept at room temperature and 100% relative humidity. The sample grid was immediately vitrified in cryogen (50/50 ethane/propane) after blotting using Vitrobot (FEI). The vitrified samples were examined under a FEI Technai G2 microscope operated at 200kV.

**Density measurements.** For the calibration plot, the densities ($\rho_{SSY}$) for the homogeneous I and N phase of pure SSY water solutions were measured using a density meter (DE45, Mettler) at room temperature. In phase separated samples, the LC region is at the bottom of a vial, while the I phase is at the top. The samples for density measurements were taken from the LC region with a syringe. The error in determining $\rho_{SSY}$ is estimated to be less than 3% on the basis of statistical analysis.

**Synchrotron x-ray studies.** For x-ray diffraction measurements, the samples were taken from the LC regions in the vials as described above and loaded into 1.5 mm diameter Lindéman capillaries with 10 μm thick walls or in sandwich-like cells made with thin (60 μm) glass plates. The samples were placed in an oven under an in-situ magnetic field of strength 2.5 kG, sufficient to align the LCLC. The sample was then exposed to synchrotron x-ray radiation of wavelength 0.7653 Å at station 6-IDB of the Midwestern Collaborative Access Team at the Advanced Photon Source of Argonne National Laboratory. The diffraction patterns were recorded at room temperature (301.4 K) using a high resolution image plate area detector, MAR345, placed at a distance of 476.0 mm from the samples. The data were calibrated against a silicon standard (NIST 640C). The intensity of the incident beam was controlled using a bank of Cu and Al attenuators. Data accumulation lasted between 1 and 60 s. The 2D diffraction patterns were analyzed using the software package FIT2D developed by Hammersley et al. [31]. The full width at half maximum of the diffraction peaks was estimated after background subtraction and used to calculate the correlation lengths.

**Spectroscopy.** SSY is a dye that absorbs in the visible and UV parts of the spectrum. To determine the concentration of SSY in coexisting phases, we used UV-vis spectroscopy of controllably diluted probes. Since the absorption coefficient might depend on SSY concentration [25], we constructed a calibration plot for concentrations from $4.5 \times 10^{-5}$ mol/l to $1.2 \times 10^{-4}$ mol/l. In this range, the maximum absorbance at the wavelength 482 nm depends linearly on the SSY concentration, as shown in the next section. The absorption spectra were measured using a spectrometer Lambda 18 (Perkin-Elmer) and rectangular cuvettes with a path length of 1cm. To determine the SSY concentrations in the I phase or N/C phase, probes were taken either from the top or bottom of the vials with well equilibrium mixtures. The probes were then diluted 5,000~16,000 times for the spectral measurements within the calibration range. Using the calibration plot and the dilution number, we determine the concentration of SSY solutions in the coexisting phases. The error in determining $c_{SSY}$ is estimated to be less than 3%. The concentration expressed in molarity units (M, the number of moles of SSY per liter of solution) is converted into the weight % units or molality unit (m, the number of moles of SSY per kilogram of water) using the following expressions: $\text{weight\%} = \frac{M \cdot W_{SSY}}{10 \rho_{SSY}}$ ; $m = \frac{1000 M}{(1000 \rho_{SSY} - M \cdot W_{SSY})}$. Here $W_{SSY}$ =452.36 g/mol is the molecular weight and $\rho_{SSY}$ is the density of SSY solution.

## 3. Experimental Results.

### 3.1. Phase diagram of ternary mixture

Figure 1a shows the phase behavior of the ternary mixture of SSY, PEG, and water at 296 K in an equilateral triangle phase diagram. Typically, in such a diagram, each vertex of the triangle represents a pure component: water, SSY, or PEG. As one moves away from the vertex, the portion of the corresponding component linearly decreases and goes to 0% at the opposite edge of the triangle. In Fig.1a, we cut the concentration range to the meaningful range of concentrations which correspond to the LC phases, so that the vertices do not correspond to pure components.

To determine the composition at any point in the diagram of Fig.1a, one uses grid lines drawn through the point of interest parallel to the edges of the triangle. For example, for the mixture labeled by *a* in Fig. 1a the composition is $c_{SSY} : c_{PEG} : c_w$ = 40:20:40 (in weight %). The left edge of the diagram shows the phase behavior of a binary SSY:water mixture, which agrees well with the previous studies [7, 25]. The homogeneous N and C phases are shown as bold lines drawn at the left edge; their thickness corresponds roughly to the extension of the *homogeneous* N and C phase upon the addition of a small amount of PEG. As $c_{SSY}$ increases, the I phase is replaced with a coexisting I+N region (26.5< $c_{SSY}$ <28.2), then a homogeneous N phase (28.2< $c_{SSY}$ <34.9), a biphasic N+C region (34.9< $c_{SSY}$ <36.1), and finally a homogeneous C region (36.1< $c_{SSY}$ <49.5). At $c_{SSY}$ >49.5, crystals start to appear in the C region. To trace the changes induced by PEG, we chose the line *1* drawn from point ***b*** ( $c_{SSY} : c_{PEG} : c_w = 28.9 : 0 : 71.1$ ) on the left edge to the pure PEG vertex. The left end point ***b*** of line *1* corresponds to a homogeneous N phase, Fig. 1b. All points on line *1* represent the same weight ratio 28.9:71.1 of SSY to water. The distance from point ***b*** increases as one increases $c_{PEG}$. Even a small amount of PEG (less than 1 wt.%) causes a transformation of the homogeneous N phase into the coexisting N and I phases. The composition of these coexisting phases is specified by the ends, ***c_N*** and ***c_I***, of the tie line passing through ***c***, Fig. 1a. The tie line is determined by the density and light absorption experiments, as described later. At $c_{PEG}$ =10wt.%, one observes three coexisting I+N+C phases (point ***d***), Fig. 1d. The ternary mixture forms the I+C biphasic state when more than $c_{PEG}$ =11wt.% is added (point ***e***), Fig. 1e.

The density ( $\rho_{SSY}$ ) of the homogeneous I and N phases of pure SSY aqueous solution increases with $c_{SSY}$, Fig. 2a. Figure 2b shows that the density ( $\rho_{LC}$ ) of the LC region condensed by the addition of PEG increases with $c_{PEG}$. Using Fig. 2, one can estimate the direction of the tie line.

The equilibrium composition of the biphasic regions is given by the intersections of a tie line with the phase boundaries. To draw the tie lines, we measured the density of the condensed LC regions, Fig. 2b. Assuming that all the PEG molecules are expelled from the condensed regions, we can use these data to estimate the SSY concentration at the left end point of the tie line, i.e., in the condensed region. The concentration of SSY in the condensed regions is higher than the SSY concentration in the solution with $c_{PEG}=0$, and increases with $c_{PEG}$, Fig. 1a. The right ends of the tie lines meet the phase boundaries of the isotropic regions, which are composed of PEG, water, and some amount of SSY.

**Table 1.** Concentration of SSY in coexisting I ($c_{SSY}^I$) and N ($c_{SSY}^N$) phases in SSY water solutions ($c_{SSY}=0.9$ mol/kg or 29 wt % ) doped with different amounts of PEG.

| $c_{PEG}$ (wt.%) | Spectroscopic measurement | | Density measurement |
|:---:|:---:|:---:|:---:|
| | $c_{SSY}^I$ (mol/kg) | $c_{SSY}^N$ (mol/kg) | $c_{SSY}^N$ (mol/kg) |
| 0.0 | - | 0.90 | 0.90 |
| 2.5 | 0.66 | 0.99 | 1.01 |
| 5 | 0.54 | 1.09 | 1.10 |
| 7.5 | 0.46 | 1.18 | 1.18 |

To determine the SSY content of the phase separated equilibrated mixtures, we also used the UV-vis spectroscopy. For strongly diluted SSY solutions, $c_{SSY}=(0.45\text{-}1.2)\times10^{-4}$ mol/l, the maximum absorbance at the wavelength of 482 nm depends linearly on $c_{SSY}$, Fig.3. Using the calibration plot in Fig. 3b, and the known degree of dilution, one can determine the SSY concentration $c_{SSY}^I$ in the I phase and its concentration $c_{SSY}^N$ (or $c_{SSY}^C$) in the coexisting N (or C) phase. Figure 4 shows that the addition of PEG to 0.9 mol/kg SSY solution decreases $c_{SSY}^I$ while increasing $c_{SSY}^N$. The UV-vis data agree well with the direct density measurements, Table 1. The PEG induced condensation of SSY into orientationally and positionally ordered phases occurs even if the initial concentration of SSY is low and corresponds to the I phase. The effect is illustrated by line *2* in Fig. 1a, where $c_{SSY} : c_w = 24.1 : 75.9$. When PEG is added at $c_{PEG} \geq 6.5$wt.%, the isotropic mixture transforms into the I+N biphasic region. However, if $c_{SSY}$ is too low, $c_{SSY} \leq 13$wt.% , adding PEG does not cause any LC condensation; see line *3* in Fig. 1a.

### 3.2. Spatial distribution of components

To explore further the partitioning of components between the coexisting phases, we used samples with 4.9wt. % of PEG and a small amount, 0.1wt.% of fluorescein isothiocyanate PEG (FITC-PEG) added to the homogeneous nematic 29wt.% SSY solution. The fluorescent FITC-PEG has been used previously to characterize the polymer-induced phase separation [32, 33]. We expect that FITC-PEG and PEG, having a similar molecular weight, behave similarly as condensing agents. Fluorescence microscope textures clearly show that in the biphasic I+N region, FITC-PEG is expelled from the ordered N phase, Fig. 5. The FITC-PEG depleted dark regions of the fluorescence image, Fig. 5a, perfectly match the SSY-rich birefringent regions of the N phase in the POM texture, Fig. 5b. One can estimate the relative concentration of FITC-PEG in both regions by comparing the fluorescent intensity, which linearly increases with the concentration of FITC-PEG when its concentration is small [34]. The fluorescence intensity in the LC region is near zero and its value in the I region is close to saturation, Fig. 5a. Using the data in Fig. 5a, we estimate that the concentration of FITC-PEG in the LC region is ~$10^4$ times smaller than that in the I region.

Cryogenic transmission electron microscopy (cryo-TEM) provides a high resolution image of the phase-separated sample. Vitrification by rapid freezing ($10^5$Ks$^{-1}$) ensures the preservation of the assembled structure and phases while avoiding possible artifacts associated with crystallization or dehydration. The cryo-TEM image, Fig.6, shows the dark N region and the bright I region in the sample with $c_{SSY} : c_{PEG} : c_w = 22.2 : 7.8 : 70.0$ (point *f* in Fig. 1a). The dark regions reveal dense collections of parallel self-assembled SSY aggregates. No long aggregates of SSY can be resolved in the bright I region. We note that in small N inclusions, smaller than $l_W \sim 0.5\,\mu m$, the SSY aggregates can orient at different angles at the N-I interface, Fig. 6. The angle α between $\hat{\mathbf{n}}$ and the normal υ to the interface can adopt all values in the range from 0º to 90º, Fig. 6. For N (and C) domains larger than $l_W \sim 0.5\,\mu m$, the surface orientation of $\hat{\mathbf{n}}$ is typically tangential, similarly to the case of DSCG [18]. The effect of different surface orientation in small domains is apparently not related to the shear alignment during TEM sample preparation as $\hat{\mathbf{n}}$ adopts different orientations within the N domains. The observation suggests that the value $l_W \sim 0.5\,\mu m$ is the lengthscale that separates two regime in domains structures, elasticity dominated at distances smaller than $l_W$, and surface-anchoring controlled at distances larger than $l_W$ [35]. For a distorted region of size $l$, the energy of elastic distortions scales as $\sim Kl$, while the surface anchoring energy scales as $\sim Wl^2$. Here $K$ is the average Frank elastic constant and $W$ is the polar anchoring coefficient. For $l < l_W$, $l_W = K/W$, the surface alignment does not follow the direction prescribed by the anisotropy of molecular interactions at the interface, as the system tends to minimize the elastic energy by avoiding director distortions. With $l_W \sim 0.5\,\mu m$, and assuming $K \sim 10\,pN$, one expects $W \sim 10^{-6}\,J/m^2$, a value close to the ones found for surface anchoring coefficients in other studies of LCLCs [10, 29, 36].

### 3.3 The correlation length of aggregates and the distance between the aggregates

Figure 7a shows the x-ray diffraction pattern of the extracted LC region of a 29wt.% SSY solution with 7.5 wt.% PEG. The x-ray patterns are the same as those obtained for the N phase of pure SSY solutions presented in Ref. [7, 27], and clearly support the H-stacking model, with the planes of the SSY molecules being perpendicular to the axis of aggregates and to $\hat{\mathbf{n}}$. Since the aromatic rings of the cores align along the magnetic field, $\hat{n}$ aligns perpendicularly to the field. We verified that this is the case of SSY phases in a separate experiment, by preparing homeotropic cells as described in Ref. [29] and applying the magnetic field normal to the cell plates, to cause director distortions. The walls of the vertical circular capillary containing the sample, being perpendicular to the field, further assist in a uniform alignment of $\hat{n}$ along the axis of the capillary. The diffraction pattern of the N phase has two pairs of arcs in the orthogonal directions, Fig 7a. One pair of arcs in the vertical direction at the large angle (2θ=13.2º) is ascribed to the stacking distance ($a_z = 0.33$ nm) between the SSY molecules in the aggregate. This broad diffraction maximum at the large angle does not move with the addition of PEG, Fig. 7c, meaning that $a_z$ is not altered by PEG. The full width at half maximum (FWHM) of the large angle peak, however, decreases in the N phase as $c_{PEG}$ increases, indicating that the correlation length $\xi_z$ of stacking measured along the aggregate axis increases with the addition of PEG, Table 2. Another pair of arcs in the horizontal direction, Fig. 7a, is from the small angle diffraction, 2θ=1.92º, which provides a measure of the average distance between the SSY aggregates. In the N phase, the diffraction line $d_1 = \lambda_{xray} / 2\sin\theta$ at the small angle shifts from 2.63 nm to 2.28 nm and the correlation lengths ($\xi_D$) determined from the FWHM value of the diffraction peaks at 2θ=1.64~1.92º increase as $c_{PEG}$ increases, Table 2. Here $\lambda_{xray}$ = 0.7653 Å is the wavelength of a synchrotron x-ray radiation.

20wt.% PEG added to the N phase of $c_{SSY}$=29wt.% SSY induces the coexistence of the C+I phases. The x-ray diffraction pattern of the condensed LC region of this mixture at small angle (2θ=2.32º) shows a strong sharp diffraction line ($d_1$) and three faint, but sharp diffraction lines ($d_2$, $d_3$, $d_4$) whose diffraction spacings are in the ratios $1:1/\sqrt{3}:1/\sqrt{4}:1/\sqrt{7}$, characteristic of the hexagonal packing of aggregates in the plane perpendicular to their axes [37, 38], Fig. 7b and c. The correlation length $\xi_D$ associated with the inter-aggregate distances, determined from the FWHM value of this diffraction peak is much larger than $\xi_D$ for the N phase, Table 2. For the hexagonal packing, the inter-aggregate axis-to-axis distance $D$ can be directly related to these diffraction lines, $D = 2d_1/\sqrt{3} = 2d_2 = 4d_3/\sqrt{3}$. Both $d_1$ and $D$ in the condensed LC regions decrease significantly as $c_{PEG}$ increases, Table 2, thus illustrating an increase of the osmotic pressure exerted by PEG onto the LC domains. The x-ray diffraction pattern of the condensed C region also shows the same large-angle peak at $2\theta=13.2°$ as the one in the N phase, indicating that the stacking distance along the aggregate's axis is not altered by PEG. The correlation length $\xi_z$, calculated from the FWHM value of this peak, is longer than $\xi_z$ for the initial PEG-free N phase, pointing to the PEG-triggered enhancement of correlated molecular stacking along the axes of aggregates. However, once the condensed region adopts the C phase order, $\xi_z$ increases to some saturated value that does not change with the further additions of PEG, Table 2.

Note that $\xi_z$ has been explicitly associated with the *correlated* packing of molecules along the aggregates rather than with the contour length $L$ of aggregates [7]. In x-ray characterization, $\xi_z$ is expected to be equal or shorter than $L$ because of the packing defects and irregularities [7]. Recent measurements by Renshaw and Day [39] of the diffusion coefficients by NMR spectroscopy that should be more sensitive to $L$ than to $\xi_z$ demonstrate that the length of aggregates is indeed larger than the correlation length measured by x-ray diffraction. Kuriabova et al. [40] arrived at a similar conclusion by performing numerical simulations of LCLCs. It would be thus of interest to explore the effect of additives by a technique sensitive to the overall length of aggregates. Such a study is certainly possible for the I phase of LCLCs (for example, by determining the diffusion coefficients as in Ref.[39]), but it might be difficult for the condensed phases. Nevertheless, an indirect evidence of elongation of the SSY aggregates in the condensed phases (besides the very appearance of the N and C phases from the I melt) comes from the fact that the SSY concentration in these phases increases with the increase of PEG amount, Table 1.

**Table 2.** X-ray diffraction date for the separated LC region of 29 wt.% SSY mixtures with PEG

| $c_{PEG}$ (wt.%) | $d_1$ (nm) | $D$ (nm) | $\xi_D$ (nm) | $\xi_z$ (nm) | Phase |
|---|---|---|---|---|---|
| 0 | 2.63 | - | 5.27 | 3.21 | N |
| 2.5 | 2.47 | - | 6.31 | 4.03 | N |
| 5 | 2.36 | - | 7.90 | 5.15 | N |
| 7.5 | 2.28 | - | 9.51 | 6.54 | N |
| 15 | 1.98 | 2.29 | 153.25 | 6.37 | C |
| 20 | 1.96 | 2.27 | 129.82 | 6.28 | C |
| 25 | 1.89 | 2.18 | 157.08 | 6.42 | C |

$$d_1 = \lambda_{xray} / 2\sin\theta$$

$D = 2d_1/\sqrt{3}$ in the C phase.

$\xi_D$ was calculated from the FWHM of a small angle peak at $2\theta=1.64 \sim 2.32°$.

$\xi_z$ was calculated from the FWHM of a large angle peak at $2\theta=13.2°$.

### 3.4 Salt effect

Since the SSY aggregates are highly charged, the phase behavior of SSY condensed by PEG can be altered by a change in the ionic strength of the solution. Our previous study [7] showed that salts added to the I phase of SSY increase the correlation length $\xi_z$ and extend the temperature region of the N phase. Jones et al. [16], however, showed an opposite trend: a large amount of NaCl ($c_{NaCl} \geq$ 1mol/L) added to a highly concentrated SSY solution resulted in a slight destabilization of the LC ordering as evidenced by the decrease of the temperatures $T_{N \to N+I}$ and $T_{N+I \to I}$. In order to clarify this apparent contradiction, we first studied the binary mixtures of SSY and water, doped with differing amounts of NaCl, namely, $c_{NaCl} = 0$; 0.5; and 1 mol/kg.

**3.4.1. Salt effects in pure SSY solutions.** Figure 8 shows that when the concentration of SSY is low, $c_{SSY} < \sim 31$wt.%, NaCl increases both $T_{N \to N+I}$ and $T_{N+I \to I}$, thus stabilizing the N phase at the expense of the I phase. However, when the concentration of SSY is high, $c_{SSY} > \sim 33$wt.%, NaCl decreases both temperatures, suppression the orientational order. The salt also suppresses the positional order pertinent to the C phase at high SSY concentrations. As seen in the right-hand part of Fig.8, NaCl added to the solution with $c_{SSY} = 37$wt.% replaces the homogeneous C phase and the coexistence C+N region with the N phase. We thus conclude that the effect of NaCl on SSY depends on the concentrations and that it can either enhance or suppress the orientational and positional order.

Previous studies showed that the polymer-induced DNA condensation is enhanced when a monovalent salt is added [22-24]. However, the non-monotonic effect of salts on SSY described above leads also to the new features of PEG-induced condensation of salted SSY that depend on whether the concentration of SSY is relatively low or high.

**3.4.2. Salt effects in SSY+PEG solutions, small $c_{SSY}$.** When $c_{SSY}$ is small, the addition of monovalent salt to the (SSY + PEG) water solution enhances the phase-separation and condensation of LC regions, similarly to the situation described for DNA. For example, the mixture of 5 wt.% PEG and 24 wt.% SSY (point *g* in Fig. 1a) that is in a homogeneous I state, Fig. 9a and b, is phase separated into the coexisting N and I phases by the addition of NaCl, Fig. 9a and c. The volume of the separated N region increases with $c_{NaCl}$, despite the fact that the total amount of SSY in all the samples shown in Fig.9 is constant.

To estimate the relative concentrations of SSY in the I and the salt-induced N region, we measured the absorbance in these two regions after a controlled dilution, as described above. According to the spectroscopic data, Fig. 9d,e, addition of NaCl decreases the concentration $c_{SSY}^{I}$ of SSY in the I phase and increases it in the N phase, from $c_{SSY}^{N} \approx 0.97$ mol/kg for $c_{NaCl} = 0.2$ mol/kg to $c_{SSY}^{N} \approx 1.09$ mol/kg for $c_{NaCl} = 1$ mol/kg, Table 3. These concentration changes in the I and N regions imply that the NaCl promotes partitioning of SSY into the condensed anisotropic regions.

**Table 3.** SSY concentration in coexisting domains of the SSY+PEG+NaCl water solutions as a function of the added salt concentration $c_{NaCl}$. Data are calculated from the absorption measurements.

| SSY+PEG | $c_{NaCl}$ (mol/kg) | $c_{SSY}^{I}$ (mol/kg) | $c_{SSY}^{N}$ (mol/kg) | $c_{SSY}^{C}$ (mol/kg) |
|---|---|---|---|---|
|  | 0 | 0.70 | - | - |
| 24% (0.7 mol/kg) SSY | 0.2 | 0.61 | 0.97 | - |
| + | 0.4 | 0.51 | 1.01 | - |
| 5%PEG | 0.6 | 0.43 | 1.05 | - |
|  | 1 | 0.36 | 1.09 | - |
| 33% (1.09 mol/kg) SSY | 0 | 0.39 | - | 1.37 |
| +  7.5%PEG | 1 | 0.24 | 1.40 | - |

**3.4.3. Salt effects in SSY+PEG solutions, high $c_{SSY}$.** When $c_{SSY}$ is high, the addition of NaCl to SSY + PEG water solutions causes an opposite effect, destabilizing the condensed LC regions. In the salt-free solution with $c_{SSY} = 33$wt.% and $c_{PEG} = 7.5$ wt.%, one observes a coexistence of the C and I phases, Fig. 10c (point **h** in Fig. 1a). However, if one adds NaCl, $c_{NaCl} = 1$ mol/kg to this sample, the C+I coexistence is changed into the N+I coexistence, Fig. 10d. The volume ratio of the I and LC domains decreases upon the addition of salt, from $V_I : V_C = 39:61$ to $V_I : V_N = 36:64$, Fig. 10a. Note that the salt-induced N phase flows much more easily than the C phase of the salt-free counterpart, Fig.10b.

Figure 10e shows that the addition of 1 mol/kg of NaCl decreases the absorbance and thus the concentration $c_{SSY}^I$ of SSY in the I phase significantly, by 38%, Table 3. The effect of NaCl on the condensed regions is different and nontrivial. First of all, the absorption and thus the SSY concentration in the condensed regions increases when NaCl is added, albeit slightly, by about 3% for $c_{NaCl} = 1$ mol/kg, Fig. 10f, Table 3. However, this increase in SSY concentration is accompanied by a replacement of the positionally-ordered hexagonal C phase with a positionally disordered N phase, Fig.10d. The combination of the C-to-N transformation and the simultaneous increase of $c_{SSY}$ in the condensed regions is highly unusual, since typically, the C-to-N transformation is associated with the decrease (rather than an increase) of the concentration of the mesomorphic units, as in Fig.1a. The absorption data suggest that the salt-induced C-to-N transformation is not a result of aggregates disintegration into shorter aggregates and monomers, as in that case $c_{SSY}^I$ and/or $V_I : V_{LC}$ would increase. As discussed in the next section, the effect can be caused by the increased flexibility of the chromonic aggregates.

## 4. Discussion.

The experimental data presented above point to the following tendencies in the phase behavior of aqueous solutions of SSY upon the addition of the neutral polymer PEG with the molecular weight 3,350 and the monovalent salt NaCl.

**(a) PEG effects.** PEG added to the I or N solutions of SSY causes a strong condensing effect on SSY, triggering phase separation and condensation of the LC phases, either of the N type or the C type. PEG causes widening of the biphasic regions, Fig.1. The concentration of SSY in the condensed regions is higher than the concentration of SSY in PEG-free solutions, Fig.4 and Table 1. UV-vis spectroscopic data demonstrates that the concentration of SSY in the I phase decreases upon the addition of PEG. The experiments with labeled PEG suggest that PEG is excluded from the condensed LC regions into the I phase.

**(b) NaCl effects.** NaCl can either enhance the orientational order and formation of the N phase in the SSY solutions, when the concentration $c_{SSY}$ is low, or suppress it, when $c_{SSY}$ is high. Moreover, when SSY solution is in the translationally ordered C phase, NaCl causes melting of this phase into the translationally disordered N phase, Fig.8.

**(c) Combined PEG and NaCl effects**. NaCl added to (SSY+PEG) solutions can either promote the phase separation and condensation of the LC phases when $c_{SSY}$ and $c_{NaCl}$ are low, or destabilize the condensed LC phases caused by PEG when the concentrations are high.

Below we describe how the excluded volume and electrostatic effects can contribute to these experimentally observed features.

(1) *Excluded volume effects: face-to-face vs side-by-side aggregation.*

PEG is widely used in biological and physical studies to condense and precipitate other molecules and particles. Partially, the reason is that PEG is capable to form numerous hydrogen bonds; each ethylene oxide monomer is expected to be associated with at least three water molecules, see for example, Ref.[41]. However, PEG causes a strong condensing effect even when its concentration is not enough to bind a substantial fraction of water. The condensing effect of PEG added in moderate quantities to a colloidal system is most often associated with the entropic "depletion attraction" between the colloidal particles first described by Asakura and Oosawa [19, 20]. The attractive force occurs when the colloidal particles are separated by distances shorter than the typical size of the polymer such as the gyration radius. Since the polymer molecules are excluded from the region between the particles, the osmotic pressure is different on the two sides of each particle; this difference results in an attractive force. Developing the concept of the depletion interaction further, Flory [42] predicted that rigid rods and flexible polymers should show very limited compatibility with each other when placed in the same solvent. The polymer promotes phase separation of rigid rods into a condensed N phase with a high concentration of rods and negligible amount of polymer, and the I phase containing few rods and practically all polymer coils. The reason for an elevated concentration of rods in the N phase is that the obstructions by neighbors for translational motion of a rod are alleviated by mutual alignment along the director. In contrast to the rods, inclusion of a globular polymer coils into the condensed N phase creates overlaps that are not mitigated by molecular orientations [42].

At the qualitative level, the SSY+PEG system does follow these general tendencies, as demonstrated by PEG-induced (a) condensation of the I phase into the N or even the C phase; (b) widening of the biphasic N+I region; (c) increase of the SSY concentration in the condensed LC regions, and decrease of SSY concentration in the coexisting I phase, Fig.4 and Table 1; (d) predominant location of PEG in the I phase, Fig.5. Note that in the Flory's model, the diameter of the polymer globules and the diameter of rigid cylindrical rods was the same [42], but the essential features are preserved when this ratio is different from 1, see, for example, Ref. [43-45].

In our experiments, the partitioning should be even more pronounced, as the gyration diameter of PEG with a molecular weight of 3,350, $d_{PEG} = 2r_g \approx 4$ nm [30], is large as compared to the diameter of SSY aggregates, $d_{SSY} \approx 1$ nm and the separation between the aggregates in the condensed phases. X-ray measurements (Table 2) show that the inter-aggregate axis-to-axis distance, $D$, estimated using the approximation $D = 2d_1/\sqrt{3}$ for the condensed phases is small, between about 3 nm and 2.2 nm. Considering the actual space between two aggregates $d_s = D - d_{SSY} \approx 1.2$-$2$ nm, one concludes that PEGs with $d_{PEG} \approx 4$ nm are too big to be intercalated in the lateral gaps between parallel rod-like aggregates of SSY in the N or C phase.

The PEG effects on the LCLCs, however, are more complex than in the systems studied previously with colloidal particles of pre-fixed shape and length [19, 20, 42-47], since the main structural unit is an assembled aggregate rather than an individual molecule; the shape of the aggregate is not fixed by covalent bonds. In LCLCs, the excluded volume effect can influence both the face-to-face stacking of molecules (i.e., the contour length $L$ of aggregates) and orientational and positional order in the side-by-side packing of aggregates. Let us approximate an LCLC "unit", an aggregate or a monomer, as a cylinder of a diameter $d_{SSY}$ and length $L = na_z$, where $n$ is the aggregation number. Each unit creates an excluded volume that is not accessible to the PEG molecules. When the two units are placed closely together, this excluded volume is reduced, which means an increase of the translational entropy of the polymer. The excluded volume effect can lead to two different geometries of depletion attraction, depending on the diameter-to-length ratio $d_{SSY}/L$. For $d_{SSY}/L \gg 1$, one expects a face-to-face stacking, while for long aggregates, $d_{SSY}/L \ll 1$, prevailing should be a side-by-side arrangement. A similar effect has been observed by Mason [48] in a different system with micron-sized disks mixed with nanometer-size spherical surfactant micelles: as the concentration of surfactant increased, the disks first assembled on top of each other into cylindrical stacks, and then these stacks formed "bundles."

To illustrate the possibility of two different geometries of the excluded-volume chromonic assembly, we approximate the PEG molecule as a sphere of radius $r_g$ and the excluded volume associated with each isolated chromonic unit as a cylinder of a diameter $d_{ssy} + 2r_g$ and a length $na_z + 2r_g$. We neglect the anisotropy of dispersive attractive forces between the SSY molecules (to be discussed later, see Eq.(4) in Sect.3). When two units stack on top of each other (face-to-face with respect to the molecular planes of SSY), the free volume available to the PEG molecules increases by a quantity that depends on the diameter of SSY and PEG molecules, but not on the aggregation number $n$:

$$V_e^{face} = \frac{\pi}{2} r_g \left( d_{SSY} + 2r_g \right)^2. \quad (1)$$

For the side-by-side placement of the same two units, with the inter-axial distance $D$ smaller than $d_{ssy} + 2r_g$, the increase in the free volume depends linearly on $n$:

$$V_e^{side} = (na_z + 2r_g) \left[ 2\left(\frac{d_{SSY}}{2} + r_g\right)^2 \cos^{-1}\left(\frac{D}{d_{SSY} + 2r_g}\right) - D\sqrt{\left(\frac{d_{SSY}}{2} + r_g\right)^2 - \left(\frac{D}{2}\right)^2} \right]. \quad (2)$$

Clearly, the excluded volume effects favor the face-to-face assembly, Fig.11a, at low values of $n$ and side-by-side assembly, Fig.11b, at higher $n$. For $d_{SSY} = 1$ nm, $a_z = 0.33$ nm, $r_g = 2$ nm, the crossover is expected at $n \approx 30$ when $D=3$ nm and at $n \approx 100$ when $D=4$ nm.

The simple geometrical argument above suggests that the excluded volume effect of PEG on SSY occurs at two different levels: for dilute system with short aggregates, the effect is mostly in the increase of the aggregate length, while for the longer aggregates in more concentrated solutions, the effect is in a denser lateral packing with an ensuing orientational and positional order, Fig.12. The first tendency correlates with the increased SSY volume fraction in the condensed regions, Fig.2 and Table 1. The second tendency has a direct experimental confirmation in the decrease of the inter-aggregate spacing as determined by x-ray diffraction, Table 2. In the next section, we discuss whether the osmotic pressure generated by PEG can be strong enough to overcome electrostatic repulsions between the SSY columns.

(2) *Osmotic pressure vs electrostatic repulsion.* PEG molecules excluded from the SSY-rich regions apply osmotic pressure on the aggregates, reducing the separation distance $D$ between them, Table 2. In the equilibrium C phase, the osmotic pressure is balanced by the screened electrostatic repulsions between the similarly charged SSY aggregates [18, 49, 50]:

$$\Pi_{PEG} = \sqrt{6\pi}\tau_{eff}^2 \frac{4k_B T}{d_{SSY}^2 l_B K_1^2(d_{SSY}/2\lambda)} \left(\frac{\lambda}{D}\right)^{\frac{3}{2}} \exp(-D/\lambda), \qquad (3)$$

where $K_1(x)$ is the first order modified Bessel function of the second type, $l_B = 0.71$ nm is the Bjerrum length at room termperature, and $\lambda$ is the decay length, equal to the Debye screening length $\lambda_D$ when the aggregates are considered as rigid rods. The Debye screening length is $\lambda_D = 1/e\sqrt{\varepsilon\varepsilon_0 k_B T/\sum_i c_i q_i^2} = 0.32$ nm for 29wt.% SSY solution, where $\varepsilon_0$ is the electric constant, $\varepsilon$ is the relative dielectric constant of water, $q$ is the ion's valency, and $e$ is the elementary charge. In the model [18, 46, 47] leading to Eq. (3), the aggregates are charged cylinders with a diameter $d_{SSY} \approx 1$ nm and "bare" dimensionless charge density $\tau = 2l_B/a_z \approx 4.2$. When the charge density is high, $\tau > 1$, $\tau$ is replaced with $\tau_{eff}$ since a certain portion of counterions bind to the aggregate surface [51]. Using Eq. (3) with $\lambda = \lambda_D = 0.32$ nm and $\tau_{eff} \approx 2.8$ [52], one estimates that for the condensed C phase with $D = 2.29$ nm (which corresponds to $c_{PEG} = 15$ wt.%), the osmotic pressure needed to overcome the lateral electrostatic repulsion of aggregates is $\Pi_{PEG} = 7.3 \times 10^5$ N/m², while for $D = 2.18$ nm ($c_{PEG} = 25$ wt.%), the needed pressure is $\Pi_{PEG} = 1.1 \times 10^6$ N/m². The actual concentrations of PEG in the I phase are higher than the values of $c_{PEG}$, and can be estimated from the relative volume of the I and the condensed LC regions. We estimated the concentration of PEG in the I phase to be in the range (22-36 wt%) when $c_{PEG}$ is in the range (15-25 wt%). For these solutions of PEG, the osmotic pressure was measured directly [53] to be in the range $(0.9 - 3.1) \times 10^6$ N/m² which correlates well with the estimates that follow from Eq. (3).

(3) *Salt effects: screening of intra-aggregate and inter-aggregate repulsions.* A universal effect of adding simple salts to the aqueous solutions of polyelectrolytes is the screening of electrostatic repulsive forces. In the case of LCLCs, these forces act within the aggregates and between the aggregates. In terms of the individual aggregate structure, the nonmonotonous effect of NaCl on the temperatures of phase transitions suggests that the screening of electrostatic repulsions by salt might lead to two opposite tendencies, one associated with the increased physical length of aggregates and another one with the decrease of their persistence length. A tentative explanation is as follows.

<u>Salt-induced elongation of aggregates.</u> Aggregation of chromonic molecules in water is promoted by their face-to-face attraction and hindered by the entropy. The balance of these two tendencies results in a polydisperse system of columns with an average aggregation number

$$\langle n \rangle \sim \sqrt{\phi} \exp \frac{E}{2k_B T}, \qquad (4)$$

determined by the volume fraction $\phi$ of the LCLC and the scission energy $E$ (the energy needed to divide an aggregate into two). Although Eq.(4) was derived for dilute isotropic solutions of uncharged molecules (see, for example, [54]), it suggests that an increase in the scission energy $E$ would promote longer aggregates and potentially an onset of orientational order. Therefore, one of the effects of salts on LCLCs might be in the effective increase of $E$ and thus $\langle n \rangle$ thanks to the screening of Coulomb repulsions between charged molecules, see Ref. [7] for more discussion. Similar effects of NaCl-triggered elongation of aggregates have been observed experimentally [55] and supported theoretically for worm-like micelles formed by surfactants [56, 57].

<u>Salt-induced reduction of persistent length.</u> Another potential result of electrostatic screening by salts is an increased flexibility of the aggregates, or shortening of their persistence length $P$. The persistence length of chromonic aggregates has not been studied yet, but we can roughly estimate the effect of salt-induced screening on $P$ following the Manning model [58] of a flexible charged rod. According to the model, charged segments of such a rod experience Coulomb repulsions, which produces a stretching force that enhances the bending rigidity of the rod. When the repulsive forces are screened by the salt, the rod becomes more flexible. The dependence of $P$ on the concentration of salt was expressed through the Debye screening length $\lambda_D$ [58]:

$$P(\lambda_D) = \frac{1}{4} \pi^{2/3} d_{SSY}^{4/3} (P^*)^{2/3} l_B^{-1} \left[ \frac{b}{\lambda_D (1 + e^{b/\lambda_D})} \left( \frac{2l_B}{b} - 1 \right) - \ln(1 - e^{-b/\lambda_D}) - 1 \right], \quad (5)$$

where $P^*$ is the persistence length of the completely neutralized "null isomer" of a charged rod and $b$ is the separation between the charges along the rod. Assuming $b = a_z/2$, we estimate that the persistence length of the SSY aggregate decreases by ~20 % if $c_{NaCl}$ increases from 0 to 1 mol/kg. Note that the decrease of $P$ upon addition of NaCl is commonly observed in water solutions of wormlike surfactant micelles [59], flexible polyelectrolytes [60], and DNA [58]. For example, for the wormlike micelles of sodium dodecyl sulfate, $P$ decreases from ~20 nm to ~10 nm when the concentration of NaCl increases from 1 mol/L to 2 mol/L [59]. For polystyrene sulfonate, $P$ was found to decrease from ~20 nm to ~5 nm when the concentration of NaCl increased from 0 mol/L to 1 mol/L [60]. Our estimates for the salt-induced decrease of $P$ for the chromonic aggregates correlate with the findings for other systems. The decrease in $P$ leads to substantial changes in the ability of aggregates to from ordered structures, as discussed below.

For flexible aggregates, the stability of LC is determined by $P$ rather than by the contour length $L$. According to the model proposed by Selinger and Bruinsma [61], as $P$ decreases, the I-to-N transition recedes to higher concentrations, in qualitative agreement with our experiments. One can expect that the non-monotonous dependence of $T_{N \to N+I}$ and $T_{N+I \to I}$ on $c_{NaCl}$ is related to two different outcomes of the salt-induced electrostatic screening, an increase of $L$ and decrease of $P$. At small $c_{NaCl}$, the aggregates elongate, thus $T_{N \to N+I}$ and $T_{N+I \to I}$ increase. At large $c_{NaCl}$, the aggregates become more flexible, which destabilizes the N phase and decreases the transition temperatures, Fig. 8. This hypothesis needs to be verified by direct measurements of the contour length $L$ and persistence length $P$ on the ionic strength.

The electrostatic screening by salt can also explain the experimentally observed melting of the C phase into the N phase. In the C phase, the salt-induced decrease of $\lambda_D$ and $P$ might result in larger undulations of aggregates and stronger fluctuations of the hexagonal lattice, causing its melting, similarly to the effects described in [56, 62].

## Conclusions

The phase behavior of the SSY solution in presence of neutral PEG and salt NaCl, described in this work, is highly nontrivial and shows different trends when the concentration of SSY is low and when these concentrations are high. The role of PEG is primarily in condensing the SSY into more densely packed regions with a higher concentration of SSY (Fig.2, Table 1) in which the system can acquire an orientational order (forming a N phase from an original I phase) or even a positional order (forming a hexagonal C phase). Since the main structural unit of LCLC is a self-assembled aggregate rather than an individual molecule, the excluded volume effects are expected to occur at two different levels, as they can lead either to (a) the face-to-face assembly of SSY molecules and the elongation of individual aggregates (when the SSY concentration is low) or (b) a tighter side-by-side packing of long aggregates when the concentration is high. The latter is clearly revealed in the decrease of the inter-aggregate separation determined by x-ray diffraction, Table 2.

The salt promotes the orientational order at low concentrations of SSY and suppresses it at high concentrations. Moreover, in highly concentrated SSY solutions, the salt is capable of melting the positionally ordered C phase into the positionally disordered N phase, regardless of the fact how the C phase has been formed (with or without the PEG). We propose that these experimental findings are caused by the salt-induced screening of electrostatic repulsion among the charged SSY molecules within and between the aggregates. The screening leads to (c) increase of the scission energy and elongation of aggregates and (d) decrease of their persistence length. The two tendencies (c) and (d) have an opposite effect on the ordered phases, as (c) promotes the stability of the N and C order, while (d) suppresses both orientational and positional order.

When PEG and NaCl are both present, they produce different combinations of the above tendencies. Low concentrations of PEG and salt enhance the onset of the N phase from weakly concentrated isotropic solutions of SSY. High concentrations of PEG promote strongly a C phase, but this tendency is mitigated when a large amount of salt is added, as the C phase melts into the N phase. All these combinations might be qualitatively categorized using the four mechanisms (a-d) described above. The rich variety of mechanism controlling the phase behavior of the chromonic lyotropic systems in presence of depletion agents and salts calls for further studies, both experimental and theoretical. On the experimental side, more structural data are needed (especially the direct measurements of the aggregate contour length and the persistent length). On the theoretical level detailed description of how the aggregate structure and their collective behavior depend on the presence of additives, both neutral and charged, remains a challenge.

**Acknowledgements**

We thank G. Tiddy and P. Collings for numerous valuable discussions. The work was supported by NSF grants DMR-076290 and MURI AFOSR FA9550-06-1-0337. Use of the Advanced Photon Source (APS) was supported by the U.S. Department of Energy (DOE), Basic Energy Sciences (BES), Office of Science, under Contract No. W-31-109-Eng-38. The Midwestern Universities Collaborative Access Team's (MUCAT) sector at the APS is supported by the U.S. DOE, BES, Office of Science, through the Ames Laboratory under Contract No. W-7405-Eng-82.

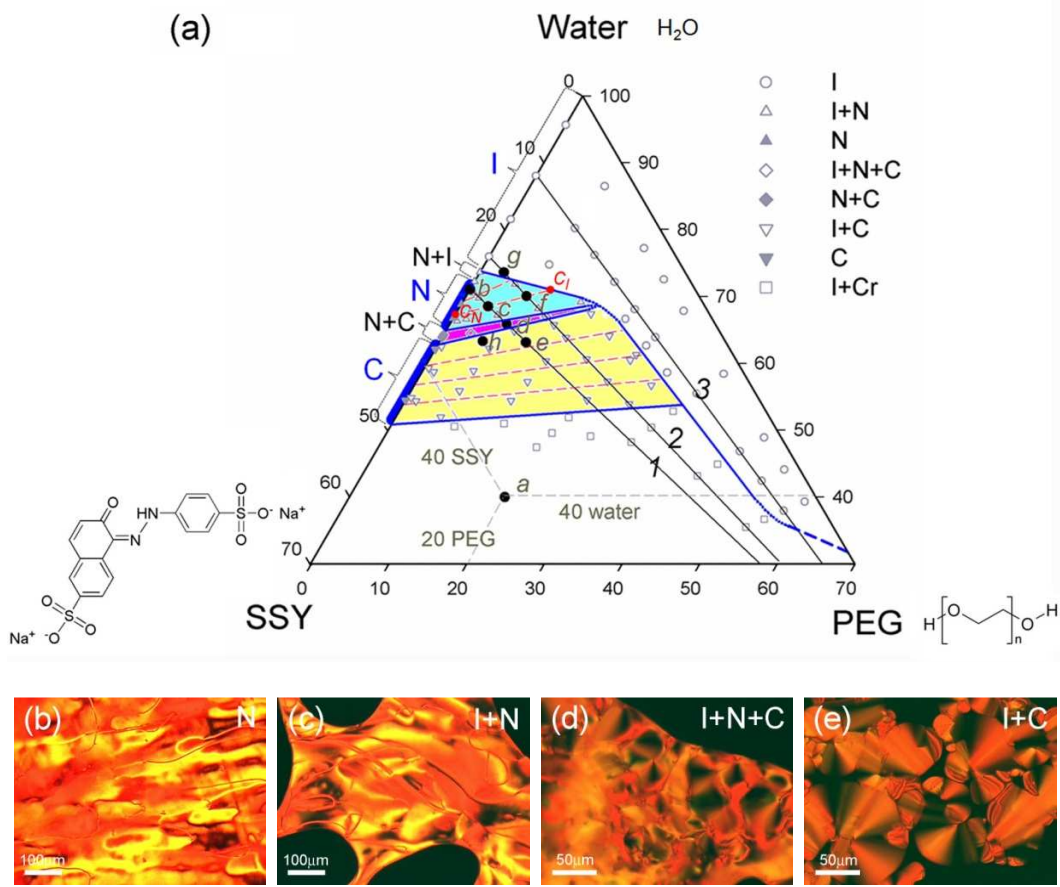

**Figure 1.** The ternary phase diagram (a) and polarizing micrographs of SSY and PEG water mixtures in the N phase (b), I+N phase (c), I+N+C phase (d), and I+C phase (e).

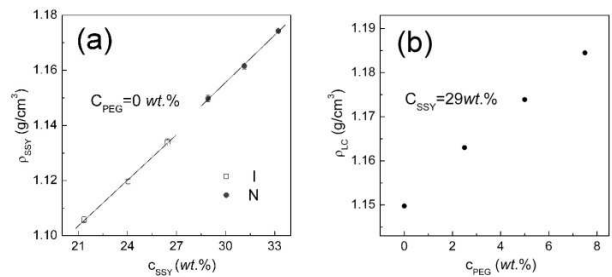

**Figure 2.** (a) Concentration dependency of density $\rho_{SSY}$ for homogeneous I and N phase of SSY solution at 296 K, (b) density $\rho_{LC}$ of the condensed LC region vs concentration of PEG.

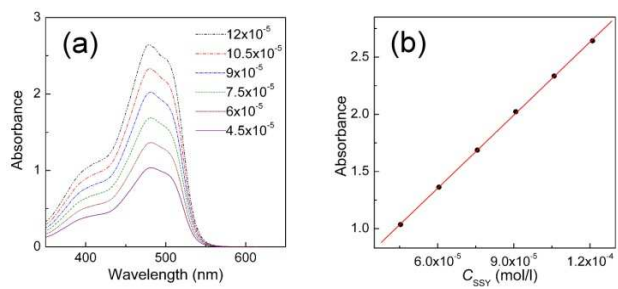

**Figure 3.** (a) Absorption spectra of pure SSY water solutions with concentrations ranging from $4.5 \times 10^{-5}$ mol/l to $1.2 \times 10^{-4}$ mol/l. (b) The linear relationship between the absorbance at $\lambda = 482$ nm and the SSY concentration.

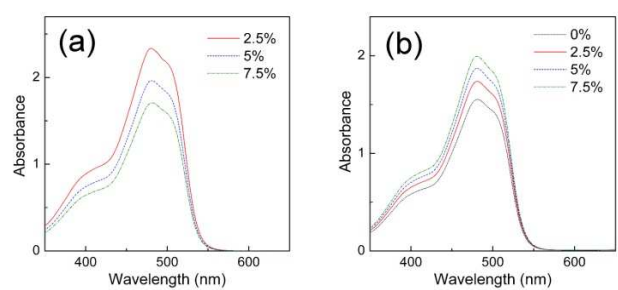

**Figure 4.** Absorption spectra of 29wt.% SSY with PEG, $c_{PEG}$ = 0wt.%; 2.5wt.%; 5wt.% and 7.5wt.% in the phase separated I (a) and N regions (b), measured for controllably diluted samples.

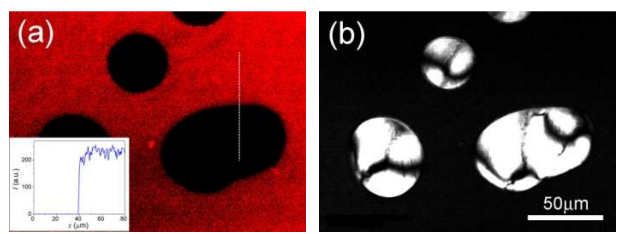

**Figure 5.** The phase separation in a 29wt.% SSY water solution caused by PEGs and FITC-PEGs. (a) Microscopic image of fluorescence and the fluorescence intensity profile along the dashed line (inset) and (b) POM texture of the same area of sample.

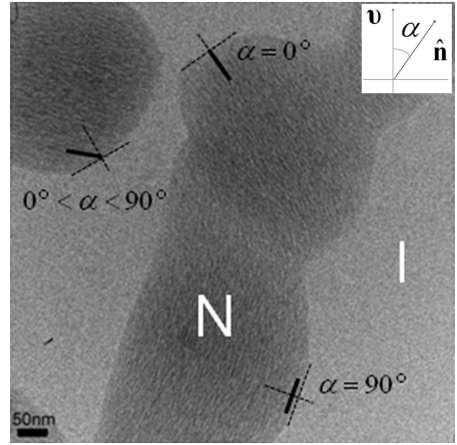

**Figure 6.** Cryo-TEM image of SSY-PEG-water mixture ( $c_{SSY} : c_{PEG} : c_w = 22.2 : 7.8 : 70.0$ ).

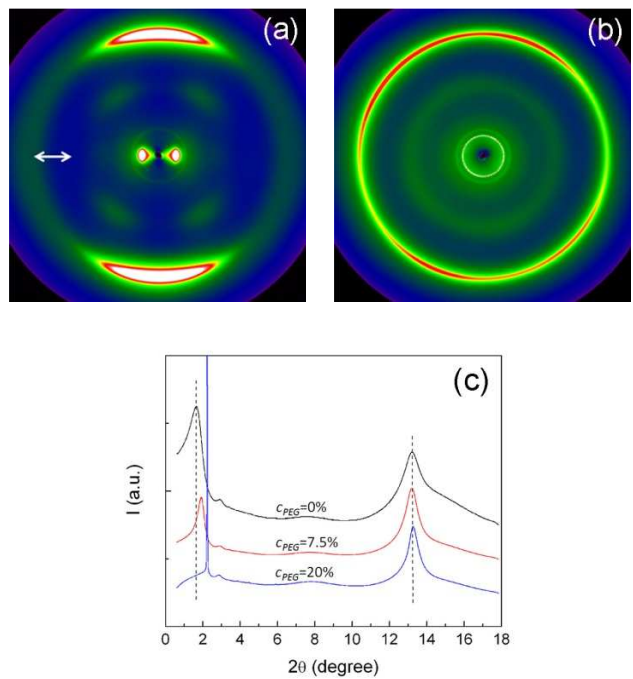

**Figure 7.** X-ray diffraction patterns of 29wt.% SSY with $c_{PEG}$ = 7.5wt.% (a) and $c_{PEG}$ = 20wt.% (b). Diffractographs of 29 wt.% SSY in the presence of PEG with $c_{PEG}$ = 0, 7.5, and 20 wt.% (c). The arrow in (a) represents the direction of the magnetic field.

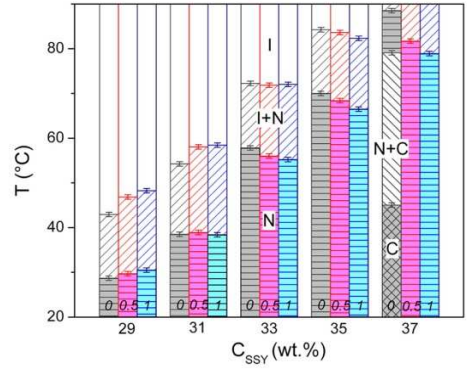

**Figure 8.** Phase behavior of SSY solution (no PEG) in the presence of NaCl, $c_{NaCl} = 0$; 0.5; and 1 mol/kg. The transition temperatures were determined upon cooling.

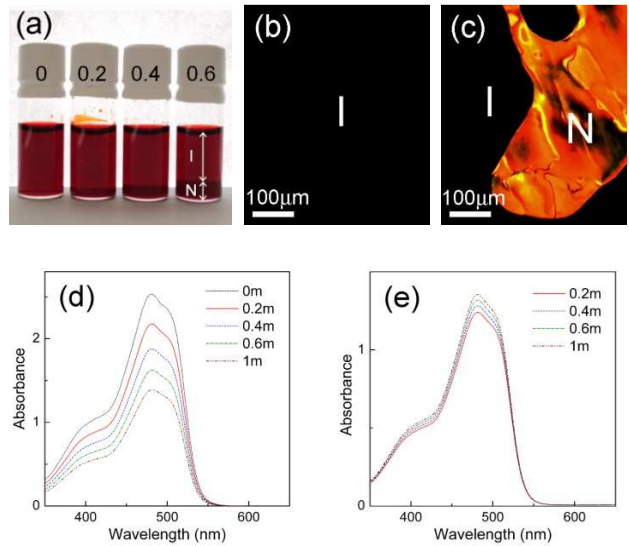

**Figure 9.** (a) Effect of NaCl on the 24 wt.% SSY + 5 wt.% PEG water solution, placed in four vials with different salt concentrations $c_{NaCl}$ indicated on the vials cups in mol/kg units. Polarizing micrographs of the I phase of the mixture $c_{SSY}:c_{PEG}:c_w = 23.1:3.9:73.0$ (b) and the N+I coexistence induced by the addition of NaCl, $c_{NaCl} = 0.2$ mol/kg to the mixture (c). Light absorption of the I phase decreases when the salt is added (d), while in the N phase it increases (e); mixture $c_{SSY}:c_{PEG}:c_w = 23.1:3.9:73.0$.

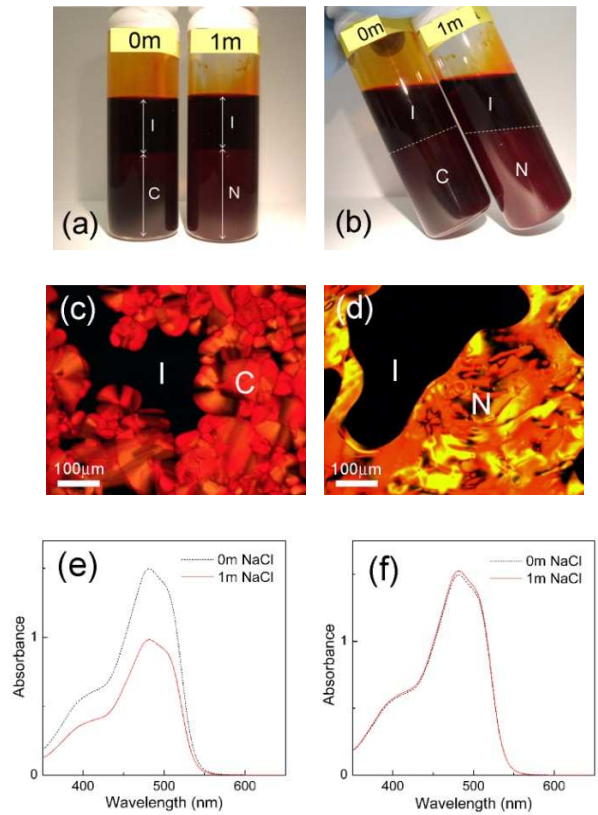

**Figure 10.** Salt effects on 33 wt.% SSY + 7.5 wt.% PEG water solution. (a) Two vials, one with $c_{NaCl}=0$ mol/kg and the C phase (left) and another one with $c_{NaCl}=1$ mol/kg (right) and the N phase, that flows easily upon tilting (b). The polarizing micrographs show the C+I coexistence for the salt-free $c_{SSY}:c_{PEG}:c_w=31.3:5.2:63.5$ mixture (c) and the N+I coexistence induced by added NaCl, $c_{NaCl}=1$ mol/kg (d). Light absorption of the I phase decreases when the salt is added (e), while in the N phase it increases (f); mixture 33 wt.% SSY + 7.5 wt.%.

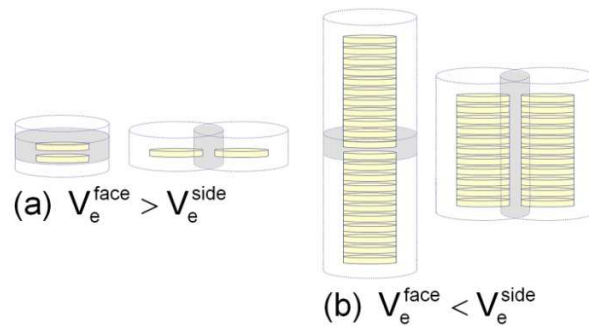

**Figure 11**. Schematic diagram shows the overlap of the excluded volumes for the face-to-face and for side-by-side placement of chromonic aggregates: (a) for individual chromonic molecules and short aggregates, face-to-face stacking releases more free volume available to other molecules such as PEG; (b) for sufficiently long aggregates, the side-by-side arrangement is favored by the excluded volume effects.

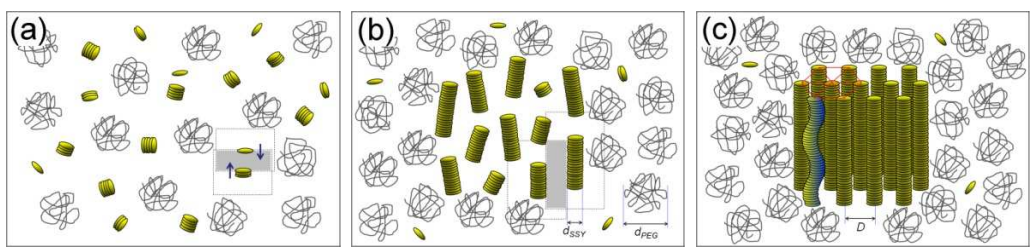

**Figure 12**. Schematic illustration of the excluded volume effect of the increasing concentration of PEG on chromonic assembly: elongation of short aggregates (a), followed by parallel arrangement in the N phase (b) and C phase (c).